\documentclass{jfm}
\usepackage{graphicx}
\usepackage{xcolor}
\usepackage{outlines}
\usepackage{amsmath}
\usepackage{verbatim}
\usepackage{epstopdf, epsfig}
\newcommand{\notetoself}[1]{\textcolor{blue}{#1}}
\newcommand{\bd}[1]{\textcolor{orange}{#1}}

\usepackage{natbib}
\usepackage{lineno}
\newcommand{\blue}[1]{\textcolor{black}{#1}}
\shorttitle{Hydraulic knots}
\shortauthor{M. V. Paludan, B. Dollet, P. Marmottant and K. H. Jensen}

\title{Elastohydrodynamic interactions in soft hydraulic knots}
\author{Magnus V. Paludan\aff{1}
	,
	Benjamin Dollet\aff{2},
	Philippe Marmottant\aff{2}
	\and Kaare H. Jensen\aff{1}\corresp{\email{khjensen@fysik.dtu.dk}}}

\affiliation{\aff{1}Department of Physics, Technical University of Denmark, Denmark
\aff{2}University Grenoble Alpes, CNRS, LIPhy, 38000 Grenoble, France}

\begin{document}

\maketitle

\begin{abstract}
%
Soft intertwined channel systems are frequently found in fluid flow networks in nature. The passage geometry of these systems can deform due to fluid flow, which can cause the relationship between flow rate and pressure drop to deviate from Hagen-Poiseuille's linear law. Although fluid-structure interactions in single deformable channels have been extensively studied, such as in Starling's resistor and its variations, the flow transport capacity of an intertwined channel with multiple self-intersections (a \textit{``hydraulic knot''}), is still an open question. We present experiments and theory on soft hydraulic knots formed by interlinked microfluidic devices comprising two intersecting channels separated by a thin elastomeric membrane. Our experiments show flow--pressure relationships similar to flow limitation, where the limiting flow rate depends on the knot configuration. To explain our observations, we develop a mathematical model based on lubrication theory coupled with tension-dominated membrane deflections that compares favorably to our experimental data. Finally, we present two potential hydraulic knot applications for microfluidic flow rectification and attenuation.
\end{abstract}


\section{Introduction}
Fluid flow in deformable channels is ubiquitous in biological and man-made fluid transportation systems. Sufficiently soft channels can be distorted by fluid pressure, yielding a pressure-dependent flow capacity that has implications on physiological flow (see, e.g., reviews by \cite{pedley2000blood} and \cite{heil2011fluid}) and is used as an enabling technology in microfluidic devices (see, e.g., the review by \cite{fallahi2019flexible}). In the vasculatures of animals and plants, interaction between fluid flow and elastic channels is critical in maintaining homeostasis, and in signal transmission through vascular networks of animals \citep{halpern1985influence,aukland1989myogenic,lautt1985mechanism} and plants \citep{louf2017universal,park2021fluid}, and to the resistance of trees to \blue{drought} \citep{choat2018triggers,keiser2022}. 

\begin{figure}
    \centering
    \includegraphics[width=\textwidth]{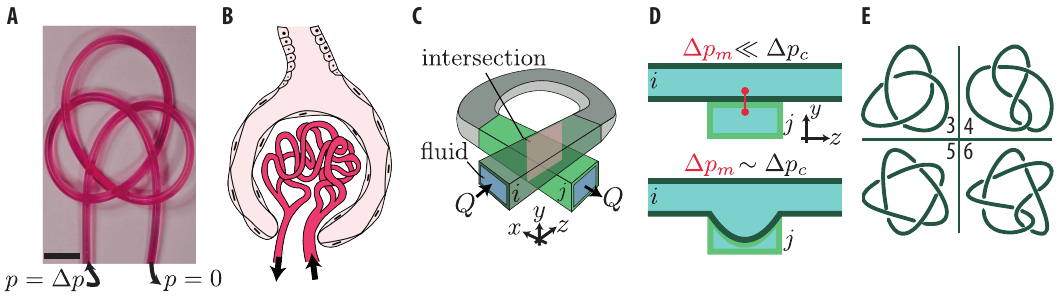}
    \caption{Fluid flow and elastic deformations can interact in self-intersecting soft channels. A) Intertwined silicone tubing filled with a red dye water solution. Pressure $p=\Delta p$ is applied to the inlet, and the outlet is connected to atmospheric conditions, $p=0$. Scale bar $= 1\,\mathrm{cm}$. B) Schematical drawing of the kidney glomerulus. The glomerulus network (red) is encapsulated in the Bowman's capsule (pink). Arrows indicate flow direction. C) Schematical drawing of a self-intersecting channel. The channel portions $(i)$ and $(j)$ are connected and overlap each other. D) Cross-sectional schematical view of the intersection (pink plane in panel C) where $(i)$ and $(j)$ intersect; $\Delta p_m$ and $\Delta p_c$ denote the transmural pressure (i.e., the fluid pressure difference between $(i)$ and $(j)$) and the characteristic elastic pressure, respectively. E) A sample of strand knots with three, four, five, and six intersections. The sketch in panel B) is adapted from www.med.libretexts.org.}
    \label{fig:Fig1}
\end{figure}

In the present work, we are interested in the transport capacity of a single microchannel that intersects itself (Fig.~\ref{fig:Fig1}A) through shared boundaries at one or more locations. A pressure source drives flow through this otherwise closed hydraulic loop. We denote this configuration a \textit{hydraulic knot}. We hypothesize that the flow--pressure characteristics for a hydraulic knot depend on the number of self-intersections and the knot's configuration. Hydraulic knots could potentially have unique hydraulic fingerprints depending on their topology, which may yield further insight into entangled physiological flows and enable new microfluidic applications. Our study on hydraulic knots builds upon the extensive previous work on fluid flow in single deformable channels, which we now briefly review.
%
%
%
%
\par\citet{knowlton1912influence} studied pressure-driven flow in a soft tube contained within a pressurized jacket to investigate blood flow autoregulation in the mammalian circulatory system. Under certain conditions, the flow rate vs. pressure relationship of the soft tube reaches a plateau in flow rate, at which point the action of increasing the applied pressure (at a specific jacket pressure) no longer yields a larger flow rate \citep{holt1941collapse,brecher1952mechanism,bertram2003experimental}. This observed \textit{flow limitation} is consistent with the myogenic response in animals' circulatory system that keeps the blood flow at an approximately steady rate \citep{klabunde2011cardiovascular}. We have recently shown that similar flow limitation patterns can occur when a long segment of a flexible conduit contacts itself, thus creating the opportunity for channel compression and the possibility of passive autoregulation \citep{Paludan2023PRE}. \blue{This system was analogous to Starling's resistor experiment (although at comparably low Reynolds number and smaller elastic deformations), where, instead of channel deformations arising from an externally applied pressure, the channel deformations arose due to the transmural pressure difference in the zone where the flexible channel self-intersects.} However, the effects of multiple intersections  (Fig.~\ref{fig:Fig1}A) have, to our knowledge, not been conducted. Many physiological flow systems, such as the kidney glomerulus capillary network (see Fig.~\ref{fig:Fig1}B), involve an intertwined densely packed conduit that intersects itself multiple times \citep{eaton2009vander}. Since the renal perfusion pressure $\Delta p \sim 10^4\,\mathrm{Pa}$ \citep{navar1978renal} is larger than the glomerular capillary wall elastic modulus $E\sim 10^3\,\mathrm{Pa}$ \citep{wyss2011biophysical}, we speculate that the soft conduit may self-compress at one or more contact points within the interlaced network, potentially leading to non-linear flow characteristics. The human umbilical cord serves as a pathway for oxygen supply and waste removal from the fetus and is a prime example of intertwined biological conduits \citep{kalish2003clinical}. The twisting of the cord provides it with turgidity, strength, and flexibility \citep{otsubo1999sonographic} while also allowing for the possibility of interaction between the vein and arteries via elastohydrodynamics because of their close proximity. Knots on the umbilical cords have been observed that can, in some cases, restrict placenta-fetus blood flow if the cord tension is large enough \citep{chasnoff1977true,sornes2000umbilical}. In the present work, we will not consider the effects of channel tension.







Biological relevant fluid-structure interactions are routinely exploited in microfluidic applications to enable micro-pumps \citep{unger2000monolithic,lee20183d}, integrated fluidic circuits \citep{thorsen2002microfluidic,leslie2009frequency,grover2006development}, and non-linear fluidic resistors \citep{gomez2017passive}. The microfluidic chips rely on stacks of flow- and auxiliary channels that intersect each other and that can be pressurized independently. Standard soft lithography and polymer molding (typically polydimethylsiloxane (PDMS) \citep{mcdonald2000fabrication}) are used to fabricate the devices, and the devices typically feature a thin PDMS membrane bonded between flow- and auxiliary channels. Because the PDMS membrane is relatively soft (Young's modulus of $E\sim 1\,\mathrm{MPa}$ \citep{lotters1997mechanical}) and thin, its deflection can be controlled via pressurizing the auxiliary channels, enabling the opportunity to close and open the flow channels selectively. Because the auxiliary- and flow channels are controlled independently, the analogy to \citeauthor{knowlton1912influence}'s externally actuated soft resistor is apparent. To our knowledge, however, comparatively little attention has been given to systems in which the flow- and auxiliary channels are connected. 


To investigate the interactions between fluids and structures in a soft intertwined channel, we will focus on a fundamental unit of a hydraulic knot, which is a soft channel that intersects itself once (see Fig.~\ref{fig:Fig1}C). Our study is limited to the simplest scenario in which rigid tissue confines the channels, thus only allowing elastic deformations in the overlap region. This situation is similar to, for example, the confinement of the afferent and efferent arteries by the Bowman's capsule in the kidney glomerulus (Fig.~\ref{fig:Fig1}B). Since the pressure decreases along the flow direction of the channel, the higher-pressure portion of the channel ($i$) can, in principle, compress the lower-pressure portion of the channel ($j$). We denote this transmural pressure drop $\Delta p_m$, arising from viscous loss in the loop that connects \textit{i} with \textit{j}. We hypothesize that when $\Delta p_m$ is sufficiently large compared to the characteristics elastic pressure $\Delta p_c$ (i.e., the pressure necessary for deforming the elastic interface), the channel portion $(i)$ can significantly deform $(j)$, thus altering the net flow capacity (Fig.~\ref{fig:Fig1}D). For an intertwined channel (such as in Fig.~\ref{fig:Fig1}A) with multiple junctions, one or several junctions can be nested inside each other, meaning that $\Delta p_m$ for one junction may depend on flow in other parts of the network. Topologically distinct knots may, therefore, have unique hydraulic signatures. \blue{This paper aims to elucidate the link between the flow rate versus pressure relationship for a hydraulic knot and its topology.}

On a broader perspective, we note that the action of intertwining a fluidic conduit is analogous to that of crafting a yarn knot, which has numerous applications in, e.g., surgery \citep{silverstein2009suturing}, fabrics \citep{warren2018clothes}, sailing \citep{mclaren2006design}, and mountaineering \citep{soles2004outdoor}. \blue{Further, fluid flow in knotted cotton yarn has applications in tunable fluidic resistance and microfluidic mixing \citep{safavieh2011microfluidics}.} Common to all these topics is the theory of knots, a rich mathematical topic \citep{adams1994knot}. The more intersections allowed, the more possible configurations exist (see a small sample of knots in Fig.~\ref{fig:Fig1}E). Whereas true mathematical knots have their strand ends connected to close the loop, our \textit{hydraulic knots} will have the inlet and outlet disconnected to allow fluid flow. Nonetheless, we will draw inspiration from the Dowker-Thistlethwaite knot notation \citep{dowker1983classification} to tabulate our hydraulic knots. 

We begin in Sec.~\ref{sec:experiment} by presenting the design, fabrication, and characterization of microfluidic PDMS devices, each comprising two perpendicularly intersecting microchannels separated by a thin PDMS membrane. We introduce our modified hydraulic knot notation in Sec.~\ref{sec:notation} and outline our experimental observations in Sec.~\ref{sec:observations_1}. To rationalize the data, we develop a mathematical model in Sec.~\ref{sec:theory_tension} inspired by \citet{Christov2018}, where the main ingredients are a tension-dominated mode of membrane deformations coupled with the low-Reynolds-number lubrication equations for the fluid flow. This allows us to predict the flow--pressure relationship for our different hydraulic knot configurations, and in Sec.~\ref{sec:comparison}, we show that the model compares favorably to our experiments. In Sec. \ref{sec:applications}, we demonstrate two applications of our microfluidic chip related to attenuating the flow rate output from a peristaltic pump (Sec.~\ref{sec:peristaltic}), and converting a purely oscillating pressure source into a net flow rate output (Sec.~\ref{sec:rectification}). Concluding remarks are given in Sec. \ref{sec:conclusion}.

\section{Experiments}\label{sec:experiment}
\begin{figure}
    \centering
    \includegraphics[width=\linewidth]{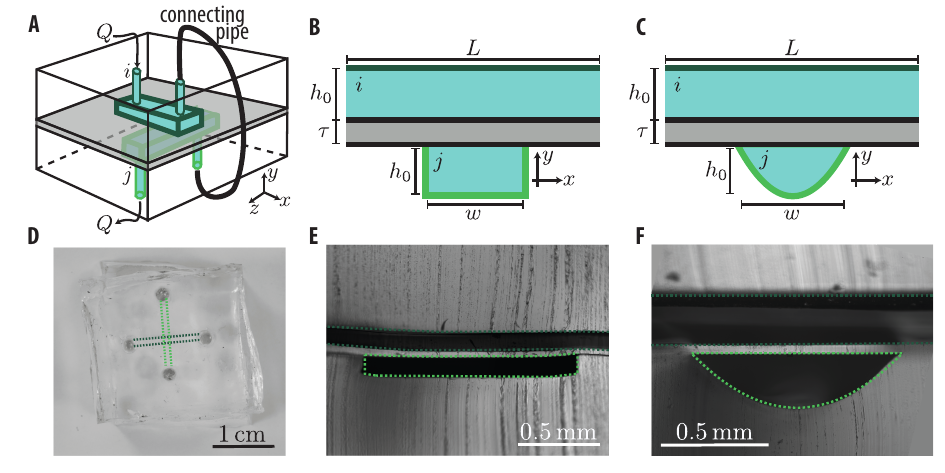}
    \caption{Microfluidic device comprising two intersecting channels. A) Schematic view of the two intersecting channels (\textit{i} and \textit{j}) separated by an elastic sheet (grey). Here, the two channels are connected via a pipe to form a single junction. B) and C) show cross-section views in the intersection when the channels are rectangular (B) and rounded (C). D) shows a top-view of the PDMS device, with the two channels overlaid with dotted lines for clarity. E) and f) show micrograph images of the channel intersections for rectangular (E) and rounded (F) channels, respectively, and the channel edges are overlaid with dotted lines for clarity. In F) the channel shape approximately follows a parabola \blue{\eqref{eq:parabola_shape}}.}
    \label{fig:Fig2}
\end{figure}

We consider a microfluidic device comprising two channels intersecting at a right angle separated by a thin elastic membrane. The two channels are arranged such that one extends in the $x-$direction, while the other extends in the $z-$direction (Fig. \ref{fig:Fig2}A and D). The device is fabricated such that the channel cross-sections are either rectangular (Fig.~\ref{fig:Fig2}B) or rounded (Fig.~\ref{fig:Fig2}C) via standard lithography techniques and PDMS molding (see details below). The PDMS membrane is clamped along the edges of the channels and has a thickness of $\tau=35\pm 5\,\mathrm{\mu m}$ as measured with a profilometer, and Young's modulus $E = 1.2\pm 0.2\,\mathrm{MPa}$ \citep{liu2009thickness}. 
The membrane can bend from the $x-z$ plane in the rectangular window where the two channels intersect. For our rectangular channels, the channel heights are $h_0 = 120\pm 10\,\mathrm{\mu m}$, widths $w = 0.96\pm 0.05\,\mathrm{mm}$, and lengths $L=1.00\pm0.05\,\mathrm{cm}$ (Fig.~\ref{fig:Fig2}B and E). For our rounded channels, the widths are $w = 0.96\pm 0.05\,\mathrm{mm}$, lengths $L=1.00\pm 0.05\,\mathrm{cm}$, and the height follows approximately the parabolic shape
\begin{align}\label{eq:parabola_shape}
    h(x) = \frac{4h_0}{w^2}\left(x-\frac{w}{2}\right)\left(x+\frac{w}{2}\right),
\end{align}
where $h_0=250\pm10\,\mathrm{\mu m}$ is the central height (Fig.~\ref{fig:Fig2}C and F). Our microfluidic device allows two configurations: either the two channels can be connected directly by an external tube (Fig.~\ref{fig:Fig1}A) in a \textit{single junction configuration}, or the channels can be connected to one or several other identical channels before looping back into the junction, in a \textit{multiple junction configurations}. In either case, the output flow rate $Q$ is measured as a function of the applied pressure $\Delta p$ (Sec.~\ref{sec:measurement_setup}).  


\subsection{Microfluidic device fabrication} \label{sec:fabrication}
Our rectangular channel devices (Figs.~\ref{fig:Fig2}B and ~\ref{fig:Fig2}E) were made by molding PDMS (Sylgard 184, Dow Chemical, MI, USA) on a patterned silicon wafer mold fabricated via standard lithography techniques. The PDMS was prepared by thoroughly mixing the base and curing agent in a 10:1 by-weight ratio and was cured on the wafer in an oven for $1\,\mathrm{hr}$ at $65^\circ\mathrm{C}$. Inlet and outlet holes were created using a biopsy punch ($2\,\mathrm{mm}$, Integra LifeSciences, NJ, USA). The membrane was made by spin-coating (WS-650-23, Laurell, PA, USA) PDMS on a clean wafer at $2500\,\mathrm{rpm}$ for $30$ seconds and was subsequently cured in the oven. The microfluidic device comprising two channels separated by the membrane was assembled by first removing the PDMS channel slabs from the molds and punching the inlet and outlet holes. One slab was then bonded to the membrane (still attached to the wafer) via plasma activation (PDC-002, Harrick Plasma, NY, USA) on high RF-power for $30\,\mathrm{s}$. After bonding, the next step was to remove the membrane (now bound to a channel) from the wafer. To do this, we first filled the channel with water using a syringe. Then, we cut around the device's perimeter with a scalpel to release the membrane from the wafer, allowing us to peel off the membrane gently. By filling the channel with water, we mitigate the risk of the membrane collapsing into the channel when the membrane is peeled off the wafer. The other channel slab was bonded to the other side of the membrane via the same procedure. Note that we aligned the channels perpendicularly by eye, although a purpose-built alignment setup (e.g., \citet{li2015desktop}) would be feasible in ensuring optimal alignment and centering.

To fabricate the rounded channel devices (Figs.~\ref{fig:Fig2}C and ~\ref{fig:Fig2}F), we followed the procedure by \citet{hongbin2009novel} to make channel molds. Briefly, the method consists of inflating a microchannel with a thin membrane lid and casting PDMS on top of the inflated membrane to yield a new channel with a height profile identical to the membrane deflection. However, instead of casting PDMS, we cast a fast-curing polymer (Elite Double 22, Zhermack, Italy) mixed 1:1 by weight. When curing was completed, we removed the polymer slab (with the channel imprint) and cast a liquid plastic resin (FormCast Burro, FormX, Netherlands) mixed 1:1 by weight to yield a rigid, rounded channel mold for our subsequent PDMS casting. This modification to \citeauthor{hongbin2009novel}'s method allows for multiple devices to be cast from the same mold without repeating the inflation step. We used an inflation pressure of \blue{$\Delta p_I = 2\,\mathrm{kPa}$} to produce our rounded channel mold, which yielded a center height of $h_0=250\pm10\,\mathrm{\mu m}$ (Fig. \ref{fig:Fig2}F). The advantage of this geometry is that it enables conforming contact between the deflected membrane and the bottom of the channel. Therefore, in this conforming geometry, the membrane can occlude the channel at a lower pressure relative to the rectangular channel geometry \citep{unger2000monolithic}, thus allowing us to study the fluid-structure interactions in multiple connected junctions within our operating pressure range. To connect device channels, we used polyvinylchloride (PVC) tubing with internal and outer diameters $1.0\,\mathrm{mm}$ and $2.0\,\mathrm{mm}$, respectively (GRA-GL0100005, Mikrolab Aarhus, Denmark). The PVC tubing's ends were cut with a razor to create a tapered tip, allowing easy insertion into the PDMS devices.

\subsection{Experimental setup}\label{sec:measurement_setup}
In characterizing our single or multiple junction devices, we measured the fluid flow rate $Q$ through the device as a function of applied pressure, $\Delta p \approx 0-40\,\mathrm{kPa}$. We used a pressure controller (LineUp Flow EZ\texttrademark, Fluigent, France) to sweep the applied pressure. The inlet of the controller was connected to a source of pressurized air, and its outlet to a closed container containing water with viscosity $\eta = (0.9\pm 0.1)\times 10^{-3}\,\mathrm{Pa\,s}$. A straw tube inserted into the container directed the pressurized water into a pressure sensor (26PC Flow-through, HoneyWell, NC, USA), and then into our microfluidic device. The device's outlet was connected to another pressure sensor and finally into a flow meter (SLF3s-1300F, Sensirion, Switzerland), connected to a reservoir held at atmospheric pressure. Using two pressure sensors kept at a constant altitude, we could accurately measure the pressure drop across the device $\Delta p$, while the flow meter provided a reading for the fluid flow rate $Q$. The pressure sensors were amplified (HX711, SparkFun Electronics, CO, USA) and connected along with the flow meter to a microcontroller (Nano Every, Arduino, Italy), and data was acquired in MATLAB (V. 2022A, MathWorks, MA, USA) using custom-built software (available upon request).

\subsection{Basic junction notation}\label{sec:notation}
\begin{figure}
    \centering
    \includegraphics[width=\linewidth]{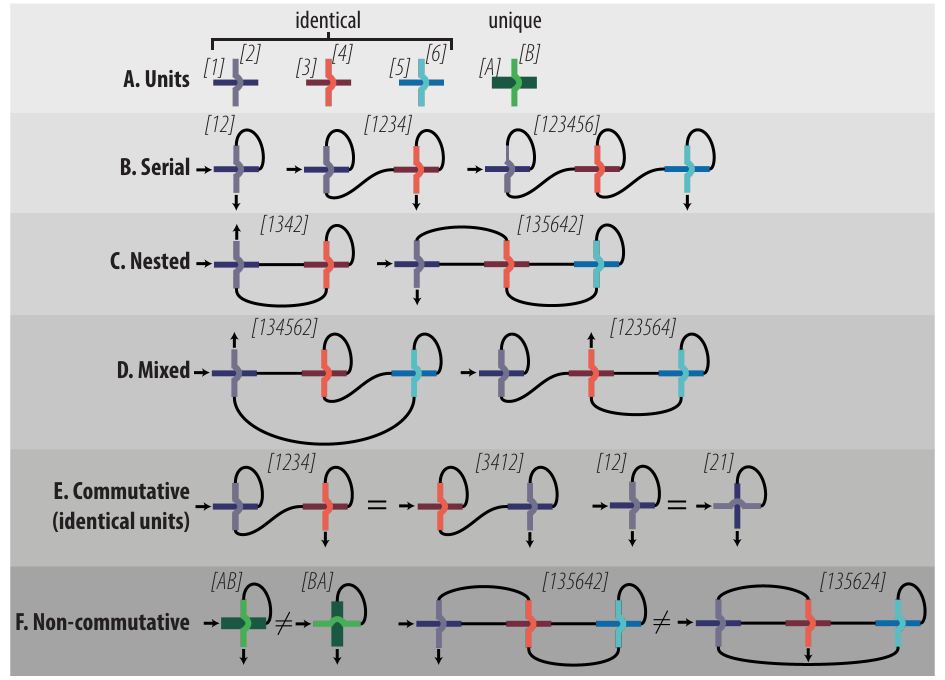}
    \caption{Tabulation of junction connections. A) Labeling of channels in identical and non-identical unit devices. Tabulation of B) serial, C) nested, and D) mixed sequences. E) Commutative rule for identical devices, and F) examples of non-commutative sequences.}
    \label{fig:Fig3}
\end{figure}

Before we proceed with experimental observations of our fluidic devices, a basic junction labeling notation must be introduced to avoid confusion about the experimental configurations. To this end, we introduce a systematic labeling technique inspired by the Dowker-Thistlethwaite notation used in the mathematical knot literature \citep{adams1994knot,dowker1983classification}. Briefly, our device encompasses two connected intersecting microchannels separated by a thin PDMS membrane. Fluid pressure causes one microchannel to deflect into the other. In our notation, we label the expanding channel an odd integer and the contracted channel an even integer. This study focuses on the hydraulic fingerprint of different sequential channel self-intersections. To this end, we will suppose that our microfluidic devices can be made with identical characteristics (i.e., identical internal channel dimensions and perfectly aligned and centered channels). For each identical device unit, we label the expanding and contracted channel, which, for three devices, yields unit one with channels \textit{[1]} and \textit{[2]}, unit two with channels \textit{[3]} and \textit{[4]}, and unit three with channels \textit{[5]} and \textit{[6]} (Fig.~\ref{fig:Fig3}A). Later (Sec.~\ref{sec:applications}), we will consider non-identical unit devices, but for now, we will label channels in a non-identical unit device letters \textit{[A]} and \textit{[B]} (Fig. \ref{fig:Fig3}A). 

Having labeled each channel, tabulating a connection between one or more unit devices is relatively straightforward, and the sequences we study can be roughly divided into three categories. The first category, \textbf{serial} sequences, is arguably the simplest and encompasses the connection of units where each unit's channels are individually connected. For instance, for unit one, the connection between channels \textit{[1]} and \textit{[2]} are made to yield the serial connection \textit{[12]} (Fig.~\ref{fig:Fig3}B). Connecting \textit{[12]} to another individually connected unit (sequence \textit{[34]}) yields another serial sequence \textit{[1234]} (Fig.~\ref{fig:Fig3}B), and similarly for the sequence \textit{[123456]} by connecting an additional unit device. However, the first channel \textit{[1]} could also be connected to \textit{[3]} instead of looping directly to \textit{[2]}. This yields the category of \textbf{nested} sequences, such as \textit{[1342]} and \textit{[135642]}, where one or more intersections are nested inside each other (Fig.~\ref{fig:Fig3}C). The third category, \textbf{mixed} sequences, encompasses serial and nested configurations that are connected to or within each other, such as \textit{[134562]} and \textit{[123564]} (Fig.~\ref{fig:Fig3}D). For identically produced unit devices, the action of replacing or switching one unit with another does not change the sequence (Fig.~\ref{fig:Fig3}E), nor does reversing the direction at which pressure is applied (e.g., by switching the sequence \textit{[12]} into \textit{[21]}, see Fig.~\ref{fig:Fig3}E)). This commutative rule is not applicable for non-identical devices (sequence \textit{[AB]}) where the flow rate depends on which direction the pressure is applied (Fig.~\ref{fig:Fig3}F). For instance, a non-identical device can be made by stacking one rounded channel \textit{[A]} and a rectangular channel \textit{[B]} in a unit device (we will explore an application related to flow rectification in Sec. \ref{sec:applications} using a non-identical device). Moreover, changing a connection in a sequence breaks the commutative rule (e.g., \textit{[135642]} differs from \textit{[135624]}, Fig.~\ref{fig:Fig3}F).  

It is worth pointing out that the number of possible sequences increases dramatically with the number of connected identical unit devices. Only one sequence, \textit{[12]}, is permitted for a single unit device, while for two unit devices, three sequences, \textit{[1234]}, \textit{[1342]}, and \textit{[1324]}, are permitted (starting with channel \textit{[1]}). For three and four unit devices, $15$ and $105$, respectively, unique sequences can be made. To keep our experiments manageable, we limit ourselves to a sample of these possible sequences representing the three sequence categories. These are listed in Table~\ref{tab:Tab1}.

\begin{table}
    \centering
    \begin{tabular}{cccc}
    Unit devices & Serial & Nested & Mixed \\ \hline 
    1 & \textit{[12]} & & \\
    2 & \textit{[1234]} & \textit{[1342]}, \textit{[1324]} & \\ 
    3 & \textit{[123456]} & \textit{[135642], [135246]} & \textit{[123564]} \\
    4 & \textit{[12345678]} & \textit{[13578642]} & \textit{[13245768]}, \textit{[12354768]}, \textit{[12357864]}\\
    \end{tabular}
    \caption{Table of hydraulic knot configurations we test. The knot notation is introduced in Sec.~\ref{sec:notation}, and our experiments sample configurations from the three categories: serial, nested, and mixed configurations.}
    \label{tab:Tab1}
\end{table}

\subsection{Observations on elementary intertwined configurations}\label{sec:observations_1}
\begin{figure}
    \centering
    \includegraphics[width=\linewidth]{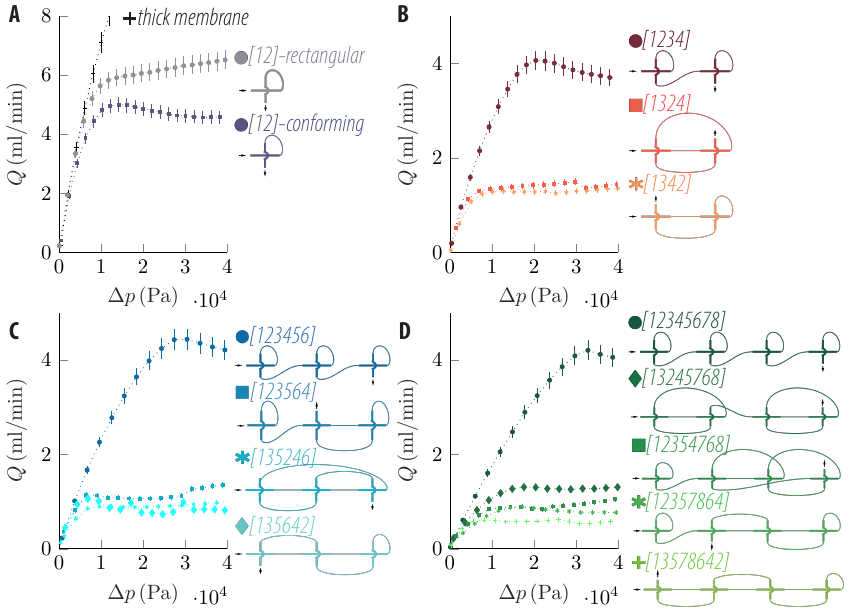}
    \caption{(Colour online) Flow rate-versus-pressure ($Q-\Delta p$) relationships for elementary hydraulic knots. The knot configurations are schematically drawn in each panel. A) The serial sequence \textit{[12]} using a single unit device with a relatively thick and thin membrane. Both conforming (purple) and rectangular (gray) junction channels were tested. In B-D, all channels have conforming cross-sections.  B), C), and D) show experimentally measured characteristics of two, three, and four unit device sequences connected according to the diagrams in the legends.}
    \label{fig:Fig4}
\end{figure}

Having outlined the experimental methods and protocol, we now focus on measured flow--pressure relationships for elementary configurations of our fluidic devices. To explore elastohydrodynamic effects in self-intersecting channels, we will first examine the simplest case: the serial connection \textit{[12]}, which involves a single unit device. To understand the impact of membrane deformations, we first measured the characteristics of the \textit{[12]} sequence with rounded channels when the membrane was relatively thick ($\tau \approx 3\,\mathrm{mm}$), inhibiting significant deformations. For this device, we found that the $Q-\Delta p$ relationship was approximately linear, in accordance with Hagen-Poiseuille's law (Fig.~\ref{fig:Fig4}A). In contrast, when the membrane was thin, a deviation from the 1:1 relationship was observed (Fig.~\ref{fig:Fig4}A). When the pressure exceeded $\Delta p\approx10\,\mathrm{kPa}$, which was necessary for significant membrane deformation, the flow rate became approximately constant for the conforming device (flow limiting regime). A deviation from the 1:1 relationship was also observed for the device with rectangular channels. However, a constant flow rate was not reached within the pressure range available in the experiment. For the conforming channel devices, we denote the flow rate plateau value by $Q_\mathrm{max}$. Below the actuation pressure, the $Q-\Delta p$ relationship was approximately linear, with a slope equal to that of the thick membrane experiment.
With two unit devices, we access the serial configuration \textit{[1234]} and two nested configurations \textit{[1324]} and \textit{[1342]}. For the serial configuration, \textit{[1234]}, we find that $Q_\mathrm{max}$ is similar to that of the \textit{[12]} sequence. However, the actuation pressure is $\Delta p\approx 20\,\mathrm{kPa}$, roughly double that of the \textit{[12]} sequence (Fig.~\ref{fig:Fig4}B). Interestingly, $Q_\mathrm{max}$ for the two nested sequences are approximately equal and lower than that of the serial sequence (Fig.~\ref{fig:Fig4}B).

Similar qualitative patterns are seen for the three and four unit device experiments (Fig.~\ref{fig:Fig4}C and D). The largest $Q_\mathrm{max}$ is found for the serial sequences, \textit{[123456]} and \textit{[12345678]}, while the lowest are found for the nested sequences, \textit{[135642]} and \textit{[13578642]}. The mixed sequences, e.g., \textit{[123564]}, have a $Q_\mathrm{max}$ that lies between that of the serial and nested sequences (Fig.~\ref{fig:Fig4}C), and similarly for the four unit device mixed sequences. A pattern also emerges for $Q_\mathrm{max}$ for nested sequences; the more nested devices, the lower the flow rate plateau, e.g., \textit{[1342]} has a larger $Q_\mathrm{max}$ than \textit{[135642]}, that again has a higher $Q_\mathrm{max}$ than \textit{[13578642]}.

We briefly summarize our experimental findings with the following qualitative statements. Serial sequences yield approximately the same $Q_\mathrm{max}$, and the pressure required for reaching $Q_{\mathrm{max}}$ scales approximately linear with the number of unit devices. Nested sequences yield the lowest $Q_\mathrm{max}$, decreasing with additional nested unit devices. Mixed sequences have a $Q_\mathrm{max}$ higher than nested and lower than serial sequences. Finally, the initial slopes of the $Q-\Delta p$ diagrams are independent of sequence configurations and depend only on the number of unit devices.

\section{Theoretical model}\label{sec:theory_tension}
We will proceed by attempting to rationalize our experimental observation on the flow limitation dependency on hydraulic knot configuration by modeling the flow-versus-pressure relationship for our fluidic devices. We start by exploring fluid-structure interactions in a single channel junction in Sec.~\ref{sec:single_junction} and the resulting flow-pressure relationship for a single knot element in Sec.~\ref{sec:single_knot}. In Sec.~\ref{sec:multiple_junctions}, we consider the coupling between multiple junctions and procedures in modeling coupled unique junctions in Sec.~\ref{sec:non-identical-geometries}. 
\subsection{Single junction}\label{sec:single_junction}
\subsubsection{General considerations and approximations}
We consider a single junction, comprising a top and a bottom channel of width $w$, crossing perpendicularly and separated by a flexible square membrane of thickness $\tau$.
As a result of the transmural pressure, the membrane bends downwards, reducing the bottom channel cross-section and increasing its hydraulic resistance.
\begin{figure}
  \centerline{\includegraphics[width=\columnwidth]{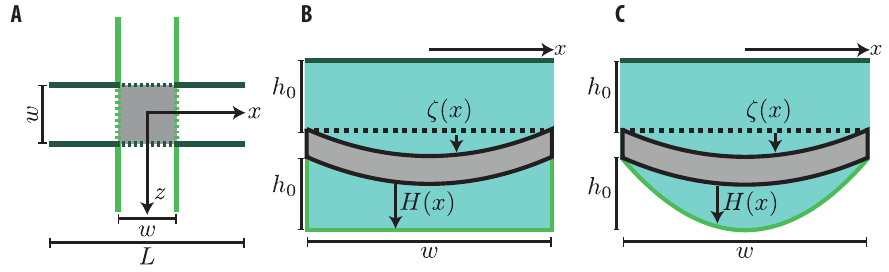}}
  \caption{(a) Top view sketch of the junction, highlighted in light shade. The top channel is in the $z$-direction, while the bottom channel is in the $x$-direction. (b) Side view sketch of the rectangular channel of width $w$ and height $h_0$. (c) Side view sketch of the conforming junction of width $w$ and center height $h_0$.}
\label{Fig:Sketch_junction}
\end{figure}
We assume that the membrane is thin, in the sense that $\tau \ll w$, and that it is clamped at its boundaries. In practice, the transmural pressure drop changes as the pressure drops along the compressed channel and is thus a function of $z$ (Fig.~\ref{Fig:Sketch_junction}a). \blue{Furthermore, the membrane is clamped on all four edges in the $w$-by-$w$ area in which the two channels intersect, which demands a complicated 2D description of the deflection in the form of cumbersome series expansions \citep{Timoshenko1987}. Even if we accounted for such solutions, solving the fluidic problem, which amounts to solving Reynolds equation $\nabla\cdot(H^3 \nabla p) = 0$ (with $H$ the channel height, see Fig.~\ref{Fig:Sketch_junction}) if the membrane slope is small \citep{bruus2007theoretical}, does not lead to analytical solutions because of the complicated form of $H$.} To keep the simplest possible approach yet retaining the constriction of the bottom channel as the essential ingredient, we ignore the dependence along the $z-$direction by assuming uniform transmural pressure $\Delta p_m$ and neglecting end effects near the clamped region at $z=\pm w/2$. Moreover, we assume that the membrane deflection $\zeta(x)$ remains small, such that $|\zeta'(x)|\ll 1$. However, the deflection is often comparable to, or larger than, the membrane thickness. Hence, the bending of the membrane induces a tension force $T$, which must be retained in the analysis \citep{Landau1986}. In such a case, the deflection obeys the equation:
\begin{align}\label{eq:FVK}
    B \frac{\mathrm{d}^4 \zeta}{\mathrm{d}x^4} - T \frac{\mathrm{d}^2 \zeta}{\mathrm{d}x^2} = \Delta p_m ,
\end{align}
where $B=E\tau ^3 /[12(1-\nu^2)]$ is the flexural rigidity (with $\nu$ the Poisson ratio), with clamped boundary conditions: 
\begin{align}\label{eq:clamped}
    \zeta = 0,\quad\frac{\mathrm{d}\zeta}{\mathrm{d}x} = 0\,\quad \mathrm{at\,}x=\pm \frac{w}{2}.
\end{align}
The solution to \eqref{eq:FVK} with boundary conditions \eqref{eq:clamped} is: 
\begin{align} \label{Eq:zeta(x)}
    \zeta(x) &= \frac{\Delta p_m w^4}{B}\left( -\frac{x^2}{2w^2\beta^2}+c_1\cosh\frac{\beta x}{w} + c_2\right),
\end{align}
where $\beta = w\sqrt{T/B}$ is a parameter comparing stretching to bending, and:
\begin{align}
    c_1 = \frac{1}{2\beta^3\sinh(\beta/2)} \quad \mathrm{and} \quad c_2 = \frac{1}{8\beta^2}-\frac{1}{2\beta^3}\coth \frac{\beta}{2} 
\end{align}
are integration coefficients. When $\beta\ll 1$, bending dominates, and the deflection $\zeta(x) \simeq \Delta p_m (x-w/2)^2(x+w/2)^2/(96B)$ follows a quartic function. When $\beta\gg 1$, stretching dominates, and the deflection $\zeta(x) \simeq \Delta p_m (w^2/4 - x^2)/(2T)$ follows a parabolic function (except for a thin boundary layer close to the clamped ends). The tension force is related to the extension of the clamped membrane upon deflection. In the limit $|\zeta'(x)|\ll 1$, it equals \citep{Landau1986}:
\begin{equation} \label{Eq:tension}
    T = \frac{E\tau}{2w} \int_{-w/2}^{w/2} \left( \frac{\mathrm{d}\zeta}{\mathrm{d}x} \right)^2 \mathrm{d}x ,
\end{equation}
which can be computed from (\ref{Eq:zeta(x)}).

We shall henceforth assume $\beta\gg 1$, an assumption which we will shortly justify. In this case, \blue{(\ref{Eq:tension})} yields $T \simeq E\tau\Delta p_m^2 w^6/(24B^2 k^4)$, whence:
\begin{align}\label{eq:k}
    \beta = 3^{1/3}\sqrt{2(1-\nu^2)}\left(\frac{\Delta p_m}{E}\right)^{1/3}\left(\frac{w}{\tau}\right)^{4/3}.
\end{align}
For our geometric and material parameters (see Sec.~\ref{sec:fabrication}), and setting $\Delta p_m = \Delta p_I = 2.0\,\mathrm{kPa}$, the parameter is $\beta \approx 15$, indicating that stretching is the dominant mode of deflection in our range of pressures. Both bending and stretching could be important to describe the deflection at relatively low pressure accurately. However, according to our data (see, e.g., Fig.~\ref{fig:Fig4}A), the flow-pressure relationships of our devices are mostly linear at relatively low pressure, indicating that the precise membrane deformation is not important at low pressure. To keep the approach simple, we, therefore, restrict the following analysis to the stretching-dominated mode of deflection, where the membrane deflection can be written as 
\begin{align}\label{Eq:deflection}
    \zeta(x) = \frac{4h_0}{w^2}\left(\frac{\Delta p_m}{\Delta p_I}\right)^{1/3}\left(\frac{w}{2} - x \right)\left(x+\frac{w}{2}\right),
\end{align}
where we have fitted the value of $\beta$ \eqref{eq:k} to the rounded channel micrograph in Fig.~\ref{fig:Fig2}F such that deflection equals the channel center height, $\zeta(0) = h_0$, when the transmural pressure equals the inflation pressure used to generate the channel shape, $\Delta p_m=\Delta p_I=2.0\,\mathrm{kPa}$. 

We denote $H(x)$ the height of the constricted bottom channel, taken independent of $z$ consistently with the aforementioned assumption: hence, the flow is unidirectional in the $z-$direction, and the pressure gradient $\nabla P$ along $z$ is uniform. \blue{Once again, this is not true in our square junctions, but it is consistent with our very simple approximation of disregarding any $z-$dependence along the junction.} We assume that the height slowly varies: $|H'(x)| \ll 1$, and that the Reynolds number is small. In this case, the relation between the flow rate $Q$ and the pressure gradient takes the following form, see, e.g., \citet{Christov2018}: 
\begin{align}\label{Eq:flow_rate_pressure_drop}
    Q = -\frac{\nabla P}{12\eta}\int_{-w/2}^{w/2} H(x)^3\,\mathrm{d}x.
\end{align}
\blue{It should be noted that the channel height $H(x)$ can also vary with the flow-direction coordinate, $z$, when the transmural pressure drop is non-uniform. However, to keep the modeling approach simple, we have neglected such dependency.} In our experiments, the effective Reynolds number \citep{Christov2018} is $\Rey'=(h_0/L)(\rho Q/(\eta w))\approx 0.2$. If the membrane deflection and the channel height were truly independent of $z$, we could replace the pressure gradient $\nabla P$ in (\ref{Eq:flow_rate_pressure_drop}) by $\Delta P_b/w$, with $\Delta P_b$ the pressure drop across the constricted bottom channel, and $w$ the length of the square junction. To account for the complicated three-dimensional structure of the membrane deflection and of the flow profile, we simply replace the relation $\nabla P = \Delta P_b/w$ by:
\begin{equation} \label{Eq:definition_Leff}
    \nabla P = \frac{\Delta P_b}{L_{\mathrm{eff}}} ,
\end{equation}
where $L_{\mathrm{eff}}$ is an effective junction length, which serves as a fitting parameter. Since the boundaries located at $z = \pm w/2$ tend to rigidify the membrane, the real deflection is overestimated by (\ref{Eq:deflection}), and the bottom channel is less constricted. Hence, we expect $L_{\mathrm{eff}}$ to be lower than $w$. Finally, inserting (\ref{Eq:definition_Leff}) in (\ref{Eq:flow_rate_pressure_drop}), the relation between the flow rate and the pressure drop across the constricted bottom channel then writes:
$$ Q = -\frac{\Delta P_b}{R_j} , $$
with the hydraulic resistance of the constricted bottom channel given by:
\begin{equation} \label{Eq:hydraulic_resistance_junction}
    R_j^{-1} = \frac{1}{12\eta L_{\mathrm{eff}}} \int_{-w/2}^{w/2} H(x)^3 \mathrm{d}x .
\end{equation}
\blue{The simplifications leading to the hydraulic resistance of the constricted bottom channel were also exploited by \citet{ozsun2013non} to estimate the nonlinear resistance of deformed channels.} We shall now calculate explicitly the flow rate--pressure drop relationship in the two cases encountered in our experiments: (i) a conforming junction, and (ii) a rectangular junction.

\subsubsection{The conforming junction}
The majority of our experiments are conducted on channels with a rounded cross-sectional shape (Fig.~\ref{fig:Fig2}F), molded from a membrane inflated by the pressure $\Delta p_I$, as discussed in Sec.~\ref{sec:fabrication}. The channel height follows approximately the parabolic shape $h(x)=4h_0/w^2(w/2-x)(w/2+x)$. As alluded to in the previous section, the compressed channel height then takes the parabolic form (Fig.~\ref{Fig:Sketch_junction}b):
\begin{align}\label{Eq:height_conforming}
    H(x) = \frac{4h_0}{w^2}\left[1-\left(\frac{\Delta p_m}{\Delta p_I}\right)^{1/3}\right]\left(\frac{w}{2}-x\right)\left(x+\frac{w}{2}\right).
\end{align}
Hence, the junction is conforming in the sense that if contact were to be established between the membrane and the bottom wall of the channel, it would extend over all the channel, thereby closing it. Inserting (\ref{Eq:height_conforming}) in (\ref{Eq:hydraulic_resistance_junction}) yields:
\begin{equation}\label{Eq:hydraulic_resistance_conforming_junction}
    R_j^{-1} = \frac{1}{R_1}\left[1-\left(\frac{\Delta p_m}{\Delta p_I}\right)^{1/3}\right]^3,
\end{equation}
where 
\begin{equation}
    R_1 = \frac{105}{4} \frac{\eta L_{\mathrm{eff}}}{wh_0^3},
\end{equation}
is the undeformed resistance of the compressed channel segment. The resistance of the straight segments of the rounded channel device, which are not influenced by membrane deflections, is given by 
\begin{align}
    R_0 = \frac{105}{4} \frac{\eta L}{wh_0^3}.
\end{align}

\subsubsection{The rectangular junction}

Some of our devices present channels with standard rectangular cross-sections of uniform $h_0$ when the membrane is undeformed (Fig.~\ref{Fig:Sketch_junction}c). When the membrane is deformed, two cases need to be analyzed, according to whether the membrane makes or not contact with the bottom wall. This has been discussed by \citet{Gilet2022} but in the case of a membrane of constant curvature. The case without contact, respectively with contact, corresponds respectively to $\zeta(0) < h_0$ and $\zeta(0) \geq h_0$, where $\zeta(0)$ is computed from (\ref{Eq:deflection}); hence, this corresponds respectively to $\Delta p_m < \Delta p_I$ and $\Delta p_m \geq \Delta p_I$.

In the absence of contact, the height profile is:
$$ H(x) = h_0 - \zeta(x) = h_0 \left[ 1 - 4 \left( \frac{\Delta p_m}{\Delta p_I} \right)^{1/3} \left( \frac{1}{2} - \frac{x}{w} \right) \left( \frac{1}{2} + \frac{x}{w} \right) \right] . $$
Inserting this expression in (\ref{Eq:hydraulic_resistance_junction}) yields:
\begin{equation} \label{Eq:flow_rate_pressure_drop_no_contact}
    R_j^{-1} = \frac{wh_0^3}{12\eta L_{\mathrm{eff}}} \left[ 1 - 2 \left( \frac{\Delta p_m}{\Delta p_I} \right)^{1/3} + \frac{8}{5} \left( \frac{\Delta p_m}{\Delta p_I} \right)^{2/3} - \frac{16}{35} \frac{\Delta p_m}{\Delta p_I} \right] .
\end{equation}

In the presence of contact, the expression (\ref{Eq:zeta(x)}) for the membrane deflection no longer holds. However, neglecting the bending term in (\ref{eq:FVK}) shows that the membrane profile remains parabolic away from the contact area. Moreover, the previous analysis showed that the curvature is given by: $\mathrm{d}^2 \zeta/\mathrm{d}x^2 = -8h_0 (\Delta p_m/\Delta p_I)^{1/3}/w^2$, see \blue{\eqref{Eq:zeta(x)}}. Hence, at given $\Delta p_m > \Delta p_I$, the portions of curved membrane outside of the contact area are parabolas of curvature $-8h_0 (\Delta p_m/\Delta p_I)^{1/3}/w^2$, tangent to the channel bottom wall at the (yet unknown) edge $x = \pm w_c/2$ of the contact area, and such that $\zeta = 0$ at $x = \pm w/2$. Straightforward algebra shows these conditions select both the contact edge location: $w_c = w[1 - (\Delta p_I/\Delta p_m)^{1/6}]$, and the following parabolic profile in the range $x\in[w_c/2,w/2]$:
$$ \zeta(x) = -\frac{4h_0}{w^2}\left(\frac{\Delta p_m}{\Delta p_I}\right)^{1/3} \left( x^2 - w_c x + \frac{1}{2} w_c w - \frac{1}{4} w^2 \right) , $$
and a symmetric profile in the range $x\in[{-w/2},{-w_c/2}]$. Expressing the height channel $H(x) = h_0 - \zeta(x)$ and inserting in (\ref{Eq:hydraulic_resistance_junction}) then yields:
\begin{equation} \label{Eq:flow_rate_pressure_drop_contact}
    R_j^{-1} = \frac{wh_0^3}{12\eta L_{\mathrm{eff}}} \times \frac{1}{7} \left( \frac{\Delta p_I}{\Delta p_m} \right)^{1/6} .
\end{equation}

The hitherto derived expressions for the hydraulic resistance of both the conforming and rectangular junction share the same form 
\begin{align}\label{Eq:hydraulic_resistance_general_form}
    R_j = R_1 f\left(\frac{\Delta p_m}{\Delta p_I}\right),
\end{align}
where $R_1$ is the undeformed resistance. The crucial point is that they are functions of the transmural pressure (through the dimensionless function $f$), which depends on the rest of the hydraulic network. We now study the entire network to quantify the impact of junction peculiarities on flow rate--pressure drop correlation. 

\subsection{The simple overlap}\label{sec:single_knot}
\begin{figure}
    \centering
    \includegraphics[width=\linewidth]{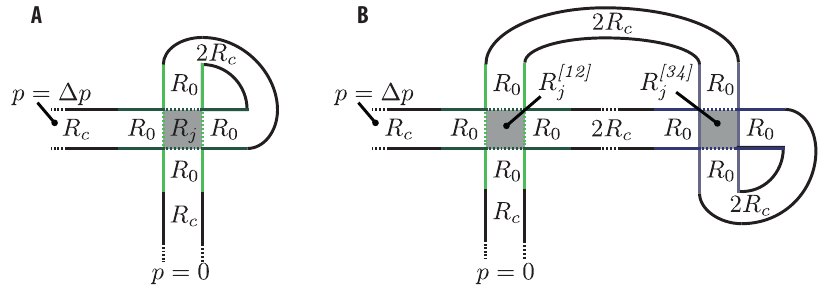}
    \caption{Schematic drawing of a) the \textit{[12]} and b) \textit{1342} junction. Straight channel segments, connective tubes, and junction resistances are termed $R_0$, $R_c$, and $R_j$, respectively. When multiple junctions are connected, e.g., in \textit{[1342]} in b), the junction resistances are labelled with superscripts according to the device channels, see Fig.~\ref{fig:Fig3}.}
    \label{fig:Fig_junctions}
\end{figure}
\subsubsection{General conditions}\label{sec:simpleoverlap_general}
We now turn to the simple network involving a junction: the simple overlap (i.e. the single serial junction \textit{[12]}). To connect channels, we fitted each straight device channel (with resistance $R_0$) with a cylindrical connecting tube. The connector has resistance $R_c = 8\eta L_c/(\pi a^4)$ (where $a$ and $L_c$ are the radius and length, respectively, with dimensions given in Sec.~\ref{sec:fabrication}). With our parameters $R_c \approx 0.5 R_0$ and thus the connecting tubes are not negligible in the total device resistance. When two channels are connected, e.g., by connecting \textit{[1]} to \textit{[2]} in the \textit{[12]} knot (see Fig.~\ref{fig:Fig_junctions}) we are left with four segments of $R_0$ resistance, four segments of $R_c$ resistance, and the junction resistance $R_j$. Following the standard additivity law of hydraulic resistors connected serially \citep{bruus2007theoretical}, the relation between applied pressure $\Delta p$ and the flow rate $Q$ is therefore 
\begin{align}
    \Delta p &= \left(4R_s + R_j\right)Q,
\end{align}
where $R_s = R_0 + R_c$ is the resistance of a straight channel segment independent of membrane deformations. Moreover, the transmural pressure across the membrane comes from the pressure drop along the path from the expanding top channel's junction outlet to the inlet of the compressed bottom junction channel, and hence it obeys $\Delta p_m = 2R_sQ$. Therefore, from \eqref{Eq:hydraulic_resistance_general_form}, $\Delta p = \left[4R_s + R_1f(\Bar{Q})\right]Q$ with a dimensionless flow rate $\Bar{Q}=Q/Q_\mathrm{max}$ with: 
\begin{align}
    Q_\mathrm{max} = \frac{\Delta p_I}{2R_s},
\end{align}
\blue{where the factor of $2$ arises from two resistive segments of $R_s = R_0 + R_c$ linking the bottom junction outlet to the top junction inlet.} We can put the characteristics in dimensionless form by setting $\overline{\Delta p} = \Delta p/\Delta p_I$ and introducing the relative resistance $\alpha = R_1/(4R_s)$, and we get: 
\begin{align}\label{Eq:DeltaP(Q)_simple_overlap}
    \overline{\Delta p} = 2\left[1+\alpha f(\Bar{Q})\right]\Bar{Q}.
\end{align}
Notice that this relationship is nonlinear because of the coupling between the flow and the membrane deformation, quantified by the function $f$. 
\subsubsection{The conforming junction}
In the case of the conforming junction, we get from \eqref{Eq:hydraulic_resistance_conforming_junction} and \eqref{Eq:hydraulic_resistance_general_form} that 
\begin{equation}\label{eq:couplingfun}
f(\Bar{Q}) = \left(1-\Bar{Q}^{1/3}\right)^{-3},
\end{equation}
which when injected in \eqref{Eq:DeltaP(Q)_simple_overlap} yields: 
\begin{equation}\label{eq:DeltaP(Q)_rounded_simple_overlap}
\overline{\Delta p} = 2\left[1+\alpha\left(1-\Bar{Q}^{1/3}\right)^{-3}\right]\Bar{Q}.
\end{equation}
The natural outcome of this calculation is the prediction of a maximal possible flow rate, equal to $Q_{\max}$. Here, flow limitation comes from the fact that at high pressure drop, any increase of the flow velocity induced by an increase of the pressure drop is exactly compensated by an increase of the hydraulic resistance of the occluding junction, leading to a saturation of the flow rate.

\subsubsection{The rectangular junction}

In the case of the rectangular junction, we get from (\ref{Eq:flow_rate_pressure_drop_no_contact}), (\ref{Eq:flow_rate_pressure_drop_contact}) and (\ref{Eq:hydraulic_resistance_general_form}) that:
\begin{align}\label{eq:f_rectangular}
     f(\bar{Q}) = \left\{ \begin{array}{ll}
\left( 1 - 2\bar{Q}^{1/3} + \frac{8}{5} \bar{Q}^{2/3} - \frac{16}{35} \bar{Q} \right)^{-1} & \mathrm{if} \, \bar{Q} \leq 1 \\
7\bar{Q}^{1/6} & \mathrm{if} \, \bar{Q} \geq 1
\end{array} \right. , 
\end{align}
which, injected in (\ref{Eq:DeltaP(Q)_simple_overlap}) yields the characteristics for rectangular channels. In marked contrast with the conforming junction, the flow rate can increase indefinitely if the pressure drop increases, according to the scaling law $\overline{\Delta p} \propto \bar{Q}^{7/6}$, which is a mild nonlinearity.

\subsection{Multiple junctions}\label{sec:multiple_junctions}
Multiple junction devices can be connected in various sequences. In our experiments, we found that the saturation flow rate $Q_\mathrm{max}$ depends on the particular sequential connection of junctions, justifying the development of a mathematical model that can predict a given sequence's $Q-\Delta p$ relationship. What makes this exercise particularly challenging is that the transmural pressure drop, $\Delta p_m$, across one channel intersection can depend on the pressure-drop-dependent resistance of other intersections. To proceed, we will start by recognizing that the hydraulic knots we study are simple serial connections of resistances. Therefore, similarly to the simple junction, we can use the additivity law of hydraulic resistors connected in series \citep{bruus2007theoretical} to write the relationship between applied pressure and output flow rate,
\begin{equation}\label{eq:flow_pressure_general_multiple_junctions}
 \Delta p = Q\left[ 4NR_s + \sum_{\text{unit devices}} R_j^{\textit{[ik]}}\left(\frac{\Delta p_m^{\textit{[ik]}}}{\Delta p_I}\right)\right],
\end{equation}
where $N$ is the number of unit devices in the knot. The first term stems from each device having four straight segments (independent of membrane deformations) of resistance $R_s$. The second term is the sum of junction resistances. Notice that we assigned the superscript $\textit{[ik]}$ to junction resistance and transmural pressure, where, consistently with Sec.~\ref{sec:notation}, the odd integer $i$ corresponds to the expanded channel, and the even integer $k$ to the contracted channel, of a given unit device. For instance, for two unit devices \textit{[12]} and \textit{[34]}, the second term in \blue{\eqref{eq:flow_pressure_general_multiple_junctions}} yields $R_j^{\textit{[12]}}+ R_j^{\textit{[34]}}$, where the junction resistances depend on the transmural pressures $\Delta p_m^{\textit{[12]}}$ and $\Delta p_m^{\textit{[34]}}$, respectively (see Fig.~\ref{fig:Fig_junctions}B). As alluded to, evaluating \blue{\eqref{eq:flow_pressure_general_multiple_junctions}} can be challenging because the junction resistances can couple. To help clarify, we shall first consider an example of a nested sequence, \textit{[1342]}.

In sequence \textit{[1342]} the junction \textit{[34]} is nested within \textit{[12]} (Fig.~\ref{fig:Fig_junctions}B), and it is clear that the transmural pressure drop of \textit{[12]}, $\Delta p_m^{\textit{[12]}}$ depends on the junction resistance of \textit{[34]}, $R_j^{\textit{[34]}}$, which depends on the transmural pressure drop of \textit{[34]}, $\Delta p_m^{\textit{[34]}}$. However, what enables us to proceed with developing the analytical model is that the transmural pressure across \textit{[34]}, $\Delta p_m^{\textit{[34]}}$, does not depend on other junction resistances. In fact, for any sequence that follows our notation, there is always at least one junction whose $\Delta p_m$ does not depend on other junctions. For the example \textit{[1342]}, this means that $\Delta p_m^{\textit{[34]}} = 2R_sQ$ (similarly to the simple overlap in Sec. \ref{sec:simpleoverlap_general}), which can be readily injected into \blue{\eqref{Eq:hydraulic_resistance_general_form}} to find the junction resistance $R_j^{\textit{[34]}}$. Knowing $R_j^{\textit{[34]}}$ enables evaluating the pressure drop across the \textit{[12]} junction as $\Delta p_m^{\textit{[12]}} = Q(6R_s+R_j^{\textit{[34]}})$, where the additional factors of $R_s$ come from the extra straight channel sections in the unit containing \textit{[34]}. Again, $\Delta p_m^{\textit{[12]}}$ can be injected into \blue{\eqref{Eq:hydraulic_resistance_general_form}} to yield the junction resistance $R_j^{\textit{[12]}}$ allowing evaluating the $\Delta p-Q$ relationship by summing all the resistive terms in \blue{\eqref{eq:flow_pressure_general_multiple_junctions}}. 

Generally, the challenging step in evaluating \blue{\eqref{eq:flow_pressure_general_multiple_junctions}} for a given sequence is identifying which junction to start with in terms of calculating the first junction-independent $\Delta p_m$ and $R_j$. However, using our knot notation, it is simply done by taking the first even integer number in a knot sequence. The next even integer junction may depend on the first or be trivial (which is true for all serial configurations). For example, in the sequence \textit{[135642]}, we recognize that \textit{[6]} is the first even number, and we calculate $\Delta p_m^{\textit{[56]}}=2R_sQ$ and $R_j^{\textit{[56]}}$ (from \blue{\eqref{Eq:hydraulic_resistance_general_form}}) which is then injected into $\Delta p_m$ and $R_j$ first for \textit{[34]} and then \textit{[12]}, finally allowing evaluating \blue{\eqref{eq:flow_pressure_general_multiple_junctions}}. In \textit{[135642]} junction \textit{[34]} depends on \textit{[56]}, and \textit{[12]} depends on both \textit{[34]} and \textit{[56]}. In contrast, the serial connection \textit{[123456]} contains only trivially solved junctions, as all of the transmural pressure drops can be written $\Delta p_m = 2R_sQ$.

We end this section by discussing the maximum attained flow rate in our hydraulic knot configurations. As was evident from our data (Fig.~\ref{fig:Fig3}), nested sequences yield a lower $Q_\mathrm{max}$ than serial sequences. In \blue{\eqref{eq:DeltaP(Q)_rounded_simple_overlap}} we found that, for a single rounded junction, the maximum flow rate $Q_\text{max} = \Delta p_I/(2R_s)$ is set by the inflation pressure (for our experiments $\Delta p_I = 2.0\,\mathrm{kPa}$), as a result of the dimensionless flow-structure coupling function $f(\Bar{Q})$. The implications on $Q_\mathrm{max}$ in our knots are best appreciated by returning to the example of the \textit{[1342]} sequence. The junction resistance $R_j^{\textit{[34]}}$ is evaluated by inserting $\Delta p_m^{\textit{[34]}} = 2R_sQ$ into \blue{\eqref{Eq:hydraulic_resistance_general_form}}, yielding 
\begin{equation}
    R_j^{\textit{(34)}} = R_1 \left[1-\left(\frac{2R_sQ}{\Delta p_I}\right)^{1/3}\right]^{-3},
\end{equation}
which predicts the maximum flow rate $Q_\mathrm{max} = \Delta p_I/(2R_s)$. To evaluate $R_j^{\textit{[12]}}$, we insert $\Delta p_m^{\textit{[12]}}=Q(6R_s+R_j^{\textit{[34]}})$ into \blue{\eqref{Eq:hydraulic_resistance_general_form}}, which gives
\begin{align}
    R_j^{\textit{[12]}} &= R_1\left[1-\left(\frac{Q(6R_s+R_j^{\textit{[34]}}}{\Delta p_I}\right)^{1/3}\right]^{-3} \nonumber \\
    &= R_1 \left\{1-\left(\frac{Q}{\Delta p_I}\right)^{1/3}\left[6R_s+R_1\left(1-\left(\frac{2R_sQ}{\Delta p_I}\right)^{1/3}\right)^{-3} \right] \right\},
\end{align}
where the reduced maximum flow rate $Q_\mathrm{max}'$ can be found by equating the braced term to zero and solving for $Q$. For the geometric and material parameters used in our experiments, $Q_\mathrm{max}' \approx 0.24 Q_\mathrm{max}$, which is in reasonable agreement with our data comparing the maximum flow rates of \textit{[12]} to \textit{[1342]} in Fig.~\ref{fig:Fig4}A and B. 

Before moving on with further comparing our mathematical model to our experimental data, we will briefly discuss modeling approaches for non-identical unit devices connected in a hydraulic knot. 

\subsection{Considerations on multiple junctions with non-identical geometries}\label{sec:non-identical-geometries}
In the previous section, we discussed modeling multiple connected identical junctions. In general, the junctions need not be identical and may differ in, e.g., channel height (linked to the inflation pressure for conforming channels), width, or length. This can result in device-specific resistances, inflation pressures, and different flow-structure coupling functions (i.e., for different channel geometries). For $N$ unit devices, the $\Delta p -Q$ relationship in \blue{\eqref{eq:flow_pressure_general_multiple_junctions}} can be adapted to 
\begin{align}
    \Delta p &= Q\left[\sum_{j=1}^{2N} R_s^{\textit{[j]}} + \sum_{\text{unit devices}} R_j^{\textit{[ik]}}
    \left(\frac{\Delta p_m^{\textit{[ik]}}}{\Delta p_I^{\textit{[ik]}}}\right)\right],
\end{align} 
where the first term sums the (different) straight channel resistances $R_s^{\textit{[j]}}$ and the second term sums the junction resistances. We will not go further into details on modeling multiple non-identical junctions, but note that we have an experiment in Sec. \ref{sec:applications} where we explore the flow-rectification features of a device that has two different channel geometries.  

\section{Comparison between experiments and theory}\label{sec:comparison}
\begin{figure}
    \centering
    \includegraphics{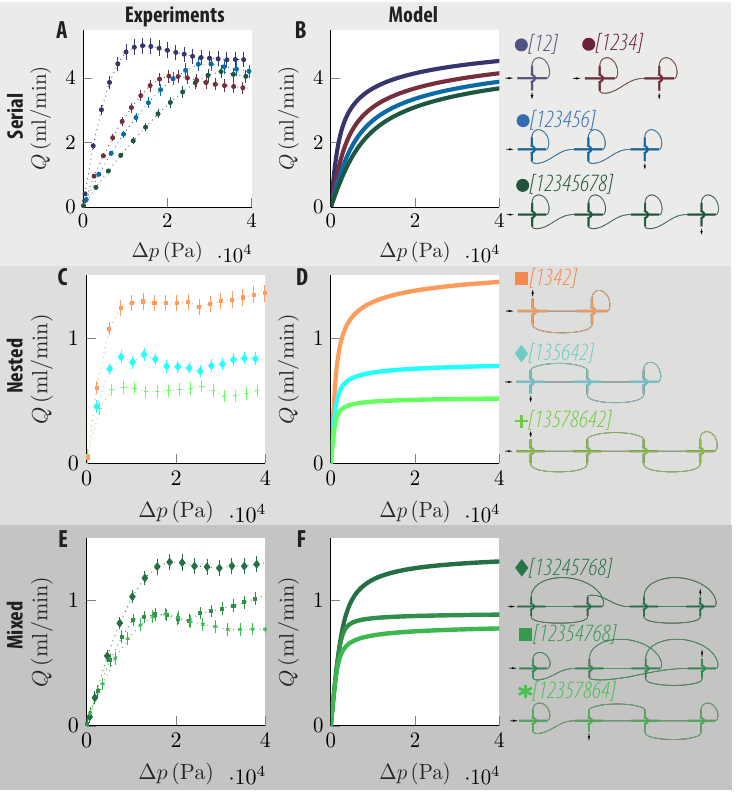}
    \caption{Comparison of our experimental data (left column) to the developed mathematical model (right column). We show experimental data and model results for serial (A and B), nested (C and D), and mixed (E and F) configurations. In each category of knots, the knots corresponding to the shown data are drawn as a legend with colors corresponding to the data colors in the plots.}
    \label{fig:Fig6}
\end{figure}

To compare our flow-structure interaction model to our experiments, we solved \blue{\eqref{eq:flow_pressure_general_multiple_junctions}} for each of our hydraulic knot configurations (Table~\ref{tab:Tab1}). To do this, we used the geometric and material parameters listed in Sec.~\ref{sec:experiment}. However, the effective overlap length $L_\mathrm{eff}$ remains undetermined. We fit $L_\mathrm{eff}\approx 0.35\,\mathrm{mm}$ to the measured flow-versus-pressure characteristics of a single knot element \textit{[12]} by minimizing the difference between the experimental and modeled maximum flow rate (at $\Delta p = 40\,\mathrm{kPa}$). \blue{Previous research has also fitted Young's modulus of PDMS from $Q-\Delta p$ data for similar hydraulic setups \citep{anand2020hydrodynamic,guyard2022elastohydrodynamic}.} Evaluating \blue{\eqref{eq:flow_pressure_general_multiple_junctions}} allows a side-by-side comparison between our experiment and model (Fig.~\ref{fig:Fig6}). For our serially connected knots, we see a qualitative and almost quantitative agreement between theory and experiment (Fig.~\ref{fig:Fig6}A and B). The onsets of flow limitation (where a digression from the 1:1 relationship becomes apparent) happen at lower pressures in the model, and the model converges slower to constant flow rates compared to experiments. In the model, the initial slope of the $Q-\Delta p$ relationship decreases for the increased number of serially connected knots, which is consistent with experiments. For our nested and mixed knots, Fig.~\ref{fig:Fig6}C and D (nested) and Fig.~\ref{fig:Fig6}E and F (mixed), the qualitative and almost quantitative agreement between model and experiment remains. Although the onset of flow limitation happens at consistently lower pressures in the model compared to experiments, the model accurately predicts the flow rate limitation levels measured experimentally. 
\begin{figure}
    \centering
    \includegraphics{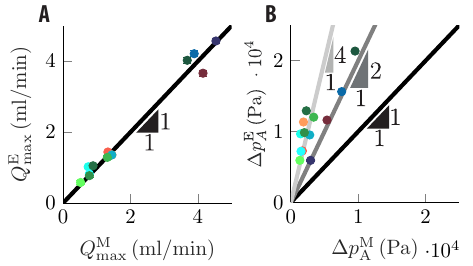}
    \caption{Comparison between A) the flow rate attained in experiments and model ($Q_\mathrm{max}^{\mathrm{E}}$ and $Q_\mathrm{max}^{\mathrm{T}}$, respectively) at the maximum applied pressure $\Delta p = 40\,\mathrm{kPa}$ and B) the actuation pressure in experiments and model ($\Delta p_\mathrm{A}^{\mathrm{E}}$ and $\Delta p_\mathrm{A}^{\mathrm{E}}$, respectively) where flow limitation ensue. In B), 1:2 and 1:4 lines are shown in dark and light gray, respectively.}
    \label{fig:Fig8}
\end{figure}

To directly compare the maximum flow rate attained by model and experiments for all our knot samples, we show, in Fig.~\ref{fig:Fig8}A, the maximum modeled flow rate, $Q_\mathrm{max}^{\mathrm{M}}$, taken at $\Delta p = 40\,\mathrm{kPa}$, i.e., the maximum pressure applied in the experiments, versus the experimental flow rate, $Q_\mathrm{max}^{\mathrm{E}}$, taken at the same pressure. The data points are colored according to the knot sequences; see legends in Fig.~\ref{fig:Fig4} and Fig.~\ref{fig:Fig6}. The solid black line indicates a 1:1 relationship between measured and predicted maximum flow rates. We see that the model favorably predicts the measured maximum flow rates despite the model's simplicity. In Fig.~\ref{fig:Fig8}B, we show the actuation pressure in the model, $\Delta p_A^{\mathrm{M}}$, versus experiment, $\Delta p_A^{\mathrm{E}}$, for all of our knot samples. We define the actuation pressure as the applied pressure where the onset of flow limitation happens. We find the actuation pressure, in both model and theory, by fitting a straight line to the linear regimes of our data and model (i.e., at relatively low pressure), which follows $Q=C\Delta p$ where $C$ is the slope. We then intercept the linear fit with the maximum attained flow rate (taken at $\Delta p=40\,\mathrm{kPa}$), such that $\Delta p_A = Q_\mathrm{max}/C$, where $\Delta p_A$ is the actuation pressure and $Q_\mathrm{max}$ is the maximum flow rate. We do this for our measurements and model, yielding the observed actuation pressure $\Delta p_A^{\mathrm{E}}$ and predicted actuation pressure $\Delta p_A^{\mathrm{M}}$. The solid black line in Fig.~\ref{fig:Fig8}B shows a 1:1 relationship between experimental and predicted actuation pressures. We see that the model consistently underestimates the actuation pressures. For the serial knots, the model predicts the actuation pressures a factor of $2$ lower than the observed, while for nested and mixed knots, the model predicts the actuation pressures a factor of roughly $4$ lower than the observed. 

One potential reason for the lack of quantitative agreement between predicted and observed actuation pressures could be our assumptions of uniform transmural pressure drops and one-dimensional membrane deflections. It is likely that the transmural pressure drop is not uniform for single junctions, such as the \textit{[12]} knot, and that the pressure drop along the junction may be important, if not dominating. As a result, the membrane deflection could depend on the $z$-coordinate (see Fig.~\ref{Fig:Sketch_junction}) and the resistance of the junction itself, leading to a description of the membrane deflection that is more complicated than our one-dimensional approximation. In addition, we neglect the finite length effects near the clamped edges of the membrane at $z=\pm w/2$, which would also demand a two-dimensional deflection description \blue{\citep{anand2020hydrodynamic}}. These combined effects may result in our model overestimating the membrane deflection at a given pressure and the degree to which the junction is compressed, leading to a lower onset of flow limitation predicted by the model compared to our experiments. Finally, it is worth mentioning that previous research has found PDMS to be hyperelastic at sufficiently large deformations \citep{nunes2011mechanical}, which can result in a scaling between transmural pressure and deflection magnitude that differs from the cubic relationship we used \citep{song2019pressure}. \blue{Finally, some of our data, specifically the serial configurations (Fig. 7A) and some mixed configurations (Fig. 7E), exhibit non-monotonic relationships between flow and pressure, whereas other configurations yield monotonic relationships between flow and pressure. Our model predicts a monotonic flow-pressure relationship for all configurations. Members of our team have previously explored soft channel devices with non-conforming channel cross-sections that also exhibit non-monotonic flow-pressure relationships \citep{park2021fluid,biviano2022smoothing}. Improving the description of the deformation in the knot junctions, as well as addressing the non-uniform junction pressure gradient, may result in non-monotonic relationships.}
While the predicted actuation pressure could potentially be improved by refining the membrane description (e.g., a two-dimensional description, and/or allowing a non-uniform transmural pressure), our model contains the necessary ingredients for quantitatively predicting the limiting flow rate for the hydraulic knots. Lastly, it is worth noting that the model predictions are quite sensitive to variations in geometric and material parameters. For instance, a modest $10\%$ variation in channel height results in a $\approx 25\%$ decrease in channel resistance ($R_0\sim h_0^{-3}$) and a $\approx 77\%$ increase in maximum flow rate for a single \textit{[12]} knot (since $\Delta p_I \sim h_0^3$ and $Q\sim \Delta p_I R_0^{-1}$). \blue{Further, a $10\%$ variation in PDMS’ Young’s modulus results in a $10\%$ variation in maximum flow rate (since $Q \sim \Delta p_I$ and $\Delta p_I \sim E$) but does not change the initial slope of the $Q-\Delta p$ diagram, since $E$ does not influence the base resistance of our system.}

\section{Applications to rectification of fluid flow}\label{sec:applications}
In this section, we introduce two possible applications of hydraulic knots: attenuating the flow rate output from a peristaltic pump and converting a purely oscillatory applied pressure to a net flow rate output by means of an anisotropic resistor. 

\subsection{Smoothing the flow rate output from a peristaltic pump} \label{sec:peristaltic}
A significant drawback of peristaltic pumps is the ripples in flow rate output that arise due to the rotational compression of the pump tubing. This unsteady flow can be unfavorable for, e.g., flow cytometry \citep{piyasena2014intersection} and infusion systems \citep{snijder2015flow}, and considerable care must be taken in smoothing the flow rate. We will consider the basic application setup where a peristaltic pump is connected to a device with resistance $R_D$. Introducing a compliant vessel between the pump and device can attenuate some, but not all, ripples in the peristaltic flow \citep{kang2012fluidic,kang2014bubble}. Further improvements can be made by including a compliant vessel and a passive non-linear resistance that only permits one specific flow rate within a range of pressure, akin to the flow limitation exhibited by our hydraulic knots \citep{doh2009passive,zhang2015passive,biviano2022smoothing}. However, the introduced non-linear resistances need to be carefully tuned such that the flow rate limitation level matches the application specifications. We hypothesize that we can overcome this obstacle by exploiting a feature of our hydraulic knots: combining multiple identical devices (each with the same flow limiting level) in different configurations can yield a spectrum of flow limiting levels (Fig.~\ref{fig:Fig4}). 
To test our hypothesis, we conducted an experiment that comprised a standard peristaltic pump (WPX1 with four rollers and $0.50\,\mathrm{mm}$ internal diameter tubing, Welco, Japan). 
We connected one end of the pump tubing to a water-filled reservoir and the other end to a 20-cm-long thin-walled silicone tubing (inner diameter $2.0\,\mathrm{mm}$, outer diameter $3.0\,\mathrm{mm}$, GRA-HS0200005, Mikrolab Aarhus, Denmark). We then connected the silicone tubing to one of our hydraulic knots and then a 10-cm-long rigid resistor ($0.50\,\mathrm{mm}$ internal diameter polyetheretherketone tubing, Mikrolab Aarhus, Denmark), which served as a proxy for the application device (following Hagen-Poiseuille's law, $R_D \approx 5.9\cdot 10^{10}\,\mathrm{Pa\,s/m^3}$ in our experiment). Using a flow meter, we measured the flow rate output from the resistive tube (see Sec.~\ref{sec:experiment}). The pump was controlled via a stepper drive (A4988, Allegro MicroSystems, NH, USA), and the pumping frequency was kept constant in our experiments at $f\approx 1.4\,\mathrm{Hz}$. 

To establish a baseline measurement of the peristaltic pump output, we conducted the experiment without installing any of our knots, i.e., with the pump connected to the compliant tube connected to the resistive tube. Next, we tested the effect of our non-linear resistive knots by installing the nested configurations \textit{[12]}, \textit{[1342]}, and \textit{[135642]} between the compliant tube and the resistive tube. The reason for using our nested knots was the observation that the flow rate limiting level decreases with the increased number of nested devices. We measured the flow rate output as a function of time for each configuration. To quantify the attenuation efficacy, we measure the ratio between peak-to-peak amplitude in flow rate ($A$) and average flow rate ($\Bar{Q}$). We define the smoothing efficacy $\varepsilon = 1-A/\Bar{Q}$, where, for smooth flow, $\varepsilon \simeq 1$, and for noisy flow, $\varepsilon \simeq 0$. 

\begin{figure}
    \centering
    \includegraphics[width=\columnwidth]{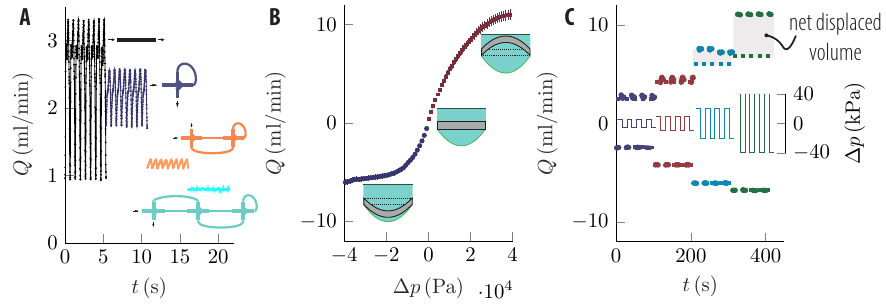}
    \caption{Soft hydraulic knots can enable microfluidics applications. A) Output flow rate from a peristaltic pump connected to a compliant tube, our hydraulic knot, and finally, a resistive tube. The black data points are a baseline with no hydraulic knots. The purple, orange, and blue data points are for the \textit{[12]}, \textit{[1342]}, and \textit{[135642]} knots, respectively. The different series are arbitrarily shifted in time for a better rendering. B) Flow-pressure characteristics of a direction-dependent resistor comprising an intersection between a rectangular and rounded channel. Schematic drawings show (not to scale) the approximate membrane position in the intersection. C) Flow-versus-time characteristics of the direction-dependent resistor when an oscillating pressure source (inset) is applied. When the pressure amplitude exceeds $\approx 20\,\mathrm{kPa}$, a net significant net volume displacement is observed during each oscillation cycle.}
    \label{fig:Fig7}
\end{figure}

For our baseline measurement, using only a compliant tube to attenuate the flow rate fluctuations, we measure a noisy flow and a low smoothing efficacy $\varepsilon \approx 0.05$ (Fig.~\ref{fig:Fig7}A). The smoothing efficacy is improved by including the \textit{[12]} knot, where we find $\varepsilon \approx 0.60$. Note that, with the inclusion of the \textit{[12]} knot, the mean flow rate remains almost the same as in the baseline case ($\Bar{Q}\approx 2.51\,\mathrm{ml/min}$ in baseline and $\Bar{Q}\approx 2.22\,\mathrm{ml/min}$ with the \textit{[12]} knot). 
The \textit{[1342]} and \textit{[135642]} knots permit lower flow rates than the \textit{[12]} knots (see, e.g., Fig.~\ref{fig:Fig4}). When we include the \textit{[1342]} knot in the peristaltic setup, the mean flow rate is reduced to $\Bar{Q}\approx 1.17\,\mathrm{ml/min}$ and the smoothing efficacy is improved to $\varepsilon \approx 0.88$. The smoothing efficacy is further improved with the \textit{[135642]} knot where, remarkably, $\varepsilon \approx 0.96$. Here, the measured flow rate oscillations are on the order of the flow rate sensor's noise level. Our experiment confirms that incorporating nested hydraulic knots into a standard peristaltic pump setup can significantly reduce pulsatile flow (to the extent of $\varepsilon \approx 0.96$) and regulate the average flow rate output by adjusting the knot sequence. For the \textit{[12]} and \textit{[1342]} knots it is not unlikely that the smoothing can be improved in response to, e.g., increasing the pumping frequency as demonstrated by \citet{biviano2022smoothing} using similar non-linear hydraulic resistances. 

\subsection{Valveless flow rectification in a direction-dependent resistor} \label{sec:rectification}
We now turn our attention to another feature of our device: the ability to intentionally mismatch the geometries of the overlapping channels. This can potentially be used to create a simple direction-dependent resistance such that the flow rate output will depend on the device's orientation (i.e., swapping the inlet and the outlet). Such diode-like fluidic resistors have been used to develop low-Reynolds-number microfluidic devices for flow rectification \citep{liu2009elastomeric,groisman2004microfluidic}. Note that, at higher Reynolds numbers, fluid inertia can be exploited to accomplish flow rectification \citep{nikola1920valvular,nguyen2021early,nguyen2021flow}.

We aim to test whether a single junction comprising one of our rounded and one of our rectangular channels can rectify flow (see Sec.~\ref{sec:experiment} for channel dimensions). We connect the rounded and rectangular channel to form a \textit{[12]} junction, and we then apply pressure to channel \textit{[1]} first (positive $\Delta p$) or channel \textit{[2]} first (negative $\Delta p$). We characterized the device similarly to our other hydraulic knots and found that the $Q-\Delta p$ characteristics are indeed asymmetric (Fig.~\ref{fig:Fig7}B). At negative $\Delta p$, the flow is limited at $Q_\mathrm{max}^{(-)}\approx -6.8 \,\mathrm{ml/min}$, while at positive $\Delta p$, the flow slowly converges to a limit at $Q_\mathrm{max}^{(+)}\approx 11.0\,\mathrm{ml/min}$. The asymmetry is caused by the membrane occluding a larger ratio of the conforming channel than the rectangular channel at a similar (absolute) applied pressure. We then performed a time-dependent experiment, using two pressure controllers to control the pressure on either end of the asymmetric device. We coded the controllers to generate a rectangular pulse of applied pressure between positive $\Delta p=(5,10,20,40)\,\mathrm{kPa}$ and negative $-\Delta p=-(5,10,20,40)\,\mathrm{kPa}$, keeping the time-averaged applied pressure $\langle\Delta p\rangle=0\,\mathrm{kPa}$ (see inset in Fig.~\ref{fig:Fig7}C). The period of each pulse was $\approx 25\,\mathrm{s}$. For small pressure pulses, $\Delta p =5-10\,\mathrm{kPa}$, the magnitude of flow rate output was approximately independent of direction. However, when we increased the pressure to $\Delta p = 20\,\mathrm{kPa}$ and $\Delta p=40\,\mathrm{kPa}$, we observed that the flow output depended on direction and that a net volume of fluid was displaced in the positive pressure direction, consistent with the measured $Q-\Delta p$ relationship in Fig.~\ref{fig:Fig7}B. We quantify the yield by the relative net flow rate output $Y=(Q^{+}-|Q^{-}|)/|Q^{-}|$, which, for our applied pressure pulses, yield $Y\approx 3\%,$ $6\%$, $21\%$, and $63\%$ in our experiment. We thus see that the yield increases with increasing applied pressure. It is not unlikely that the yield can be tuned by combining several asymmetric devices in optimized hydraulic knot configurations.    

\section{Discussion and conclusion}\label{sec:conclusion}
We have created a microfluidic device that consists of two intersecting channels separated by a thin membrane. By connecting multiple identical devices in different configurations, we studied how fluid-structure interactions affect the flow capacity of an intertwined soft channel. Our hydraulic knots exhibited flow-pressure characteristics similar to flow limitation for conforming channels, where the output flow rate remains constant and independent of further applied pressure after a critical pressure necessary for sufficient elastic deformations is reached. We categorized our devices into three configurations: serial, nested, and mixed, and measured the flow-versus-pressure relationship for a sample of these hydraulic knots. We observed three qualitative trends: serial knots have almost the same flow rate limitation level, the maximum flow rate decreases with increasing nested intersections, and the flow rate level of mixed knots lies between serial and nested knots. 

To understand our experimental observations better, we developed a mathematical model inspired by the work of \citet{Christov2018} that predicts the relationship between flow and pressure for hydraulic knots. Our model was based on low-Reynolds-number lubrication theory and tension-dominated membrane deformations, and we found an expression for the $Q-\Delta p$ relationship of single conforming and rectangular channel junctions. We then expanded the model to encompass the connection of multiple identical junctions and briefly discussed the approach to coupling multiple non-identical junctions. We found that our model compares favorably to experimental observations, and we explained the dependence of the maximum flow rate on the network topology. Notably, our model could quantitatively predict our hydraulic knots' flow rate limitation levels. However, our model consistently underestimates the actuation pressure, where the onset of flow limitation happens. 

We concluded our study by characterizing two possible microfluidics applications using hydraulic knots. Firstly, we showed that fluid flow output from a standard peristaltic pump setup could be attenuated and shifted by including different nested hydraulic knots in the hydraulic circuit downstream from the pump. Secondly, we designed a direction-dependent hydraulic resistor based on using two different channel cross-sections (conforming and rectangular) in our single hydraulic knot element. This, we showed, could be used to generate a net flow rate output from an oscillatory pressure source.

\blue{In the context of physiological blood flow autoregulation in entangled capillary bed networks, e.g., in the kidney glomerulus (Fig.~\ref{fig:Fig1}B), our study indicates that flow limitation due to fluid-structure interactions in overlapping channels is plausible. However, additional research is required before our knot model can be applied to the flow transport capacity of actual organs, specifically regarding the arrangement of intersecting vessels and the geometric and mechanical properties of the vessels and surrounding tissue.}

\section*{Acknowledgements}
M.V.P. and K.H.J were supported by Research Grant No. 9064-00069B from the Independent Research Fund Denmark. B. D. and P. M. acknowledge support from the French Agence Nationale de la Recherche (grant no. ANR619-CE30-0010-02). \blue{The authors thank the anonymous referees for their valuable feedback on the manuscript.}

\section*{Declaration of interests}
The authors report no conflict of interest.
\bibliographystyle{jfm}

\end{document}